\documentclass[12pt]{article}
\usepackage{graphicx}
\usepackage{epsfig}
\usepackage{wrapfig}
\usepackage{bm}
\usepackage{amsmath}
\usepackage{amsfonts}
\usepackage{amssymb}
\begin{document}

\title{$VA\widetilde{V}$ correlator within the instanton vacuum model}
\author{A.E. Dorokhov\\{\small Bogoliubov Laboratory of Theoretical Physics, Joint Institute for
Nuclear Research, }\\{\small 141980 Dubna, Russia}}
\maketitle
\begin{abstract}
The correlator of vector and nonsinglet axial-vector currents in the external
electromagnetic field is calculated within the instanton liquid model of QCD
vacuum. In general the correlator has two Lorentz structures: longitudinal
$w_{L}$ and transversal $w_{T}$ with respect to axial-vector index. Within the
instanton model the saturation of the anomalous $w_{L}$ structure is
demonstrated. It is known that in the chiral limit the transversal structure
$w_{T}$ is free from perturbative corrections. In this limit within the
instanton model we calculate the transversal invariant function $w_{T}$ at
arbitrary momentum transfer $q$ and show the absence of power corrections to
this structure at large $q^{2}$. Instead there arise the exponential
corrections to $w_{T}$ at large $q^{2}$ reflecting nonlocal properties of
QCD\ vacuum. The slope of $w_{T}$ at zero virtuality, the QCD vacuum magnetic
susceptibility of the quark condensate and its momentum dependence are estimated.
\end{abstract}

\section{Introduction}

Since discovery of anomalous properties \cite{Adler:1969gk,BJ} of the triangle
diagram (Fig. 1) with incoming two vector and one axial-vector currents
\cite{Rosenberg:1963pp} many new interesting results have been gained.
Recently the interest to triangle diagram has been renewed due to the problem
of accurate calculation of higher order hadronic contributions to muon
anomalous magnetic moment via the light-by-light scattering
process\footnote{See, \textit{e.g.,} \cite{APP02,CMV} and references
therein.}. At low energies the dynamics of light-by-light scattering is
nonperturbative, so one needs rather realistic QCD inspired model to find a
solution with the lowest model sensitivity.

The light-by-light scattering amplitude with one photon real and another
photon has the momenta much smaller than the other two, can be analyzed using
operator product expansion (OPE). In this special kinematics the amplitude is
factorized into the amplitude depending on the largest photon momenta and the
triangle amplitude involving the axial current $A$ and two electromagnetic
currents (one soft $\widetilde{V}$ and one virtual $V$). The corresponding
triangle amplitude, which can be viewed as a mixing between the axial and
vector currents in the external electromagnetic field, were considered
recently in \cite{CMV,VainshPLB03}. It can be expressed in terms of the two
independent invariant functions, longitudinal $w_{L}$ and transversal $w_{T}$
with respect to axial current index. In perturbative theory for massless
quarks (chiral limit) one has for space-like momenta $q$ $\left(  q^{2}%
\geq0\right)  $
\begin{equation}
w_{L}\left(  q^{2}\right)  =2w_{T}\left(  q^{2}\right)  =\frac{2}{q^{2}}.
\label{WLTch}%
\end{equation}
The appearance of the longitudinal structure is the consequence of the axial
Adler-Bell-Jackiw anomaly \cite{Adler:1969gk,BJ}. Because there are no
perturbative (Fig. 1b) \cite{Adler:er} and nonperturbative (Fig. 1c)
\cite{tHooft} corrections to the axial anomaly, the invariant function $w_{L}$
remains intact when interaction with gluons is taken into account.
Nonrenormalization of the longitudinal part follows from the 't~Hooft
consistency condition \cite{tHooft}, i.e.\ the exact quark-hadron duality. In
QCD this duality is realized as a correspondence between the infrared
singularity of the quark triangle and the massless pion pole in terms of
hadrons. It was shown in \cite{VainshPLB03} (see also \cite{Knecht04}) that in
nonsinglet channel the transversal structure $w_{T}$ is also free from
perturbative corrections. OPE\ analysis indicates that at large $q$ the
leading nonperturbative power corrections to $w_{T}$ can only appear starting
with terms $\sim1/q^{6}$ containing the matrix elements of the operators of
dimension six \cite{Knecht02}. Thus, the transversal part of the triangle with
a soft momentum in one of the vector currents has no perturbative corrections
nevertheless it is modified nonperturbatively.

\begin{figure}[th]
\includegraphics[width=14cm]{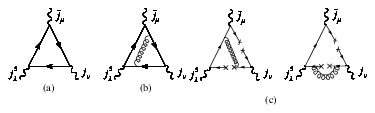}\caption{Quark triangle diagram, $(a)$;
perturbative gluon, $(b),$ and four-quark condensate, $(c),$ corrections to
it.}%
\label{fig:triangle}%
\end{figure}

In the present work we analyze in the framework of the instanton liquid model
\cite{ShSh} the nonperturbative properties of the triangle diagram in the
kinematics specified above (see Section 2 for further details). The model is
based on the representation of QCD\ vacuum as an ensemble of strong vacuum
fluctuations of gluon field, instantons. They characterize nonlocal properties
of QCD\ vacuum \cite{MikhRad92,DEM97,DoLT98}. The interaction of light $u,d$
quarks in the instanton vacuum can be described in terms of effective 't Hooft
four-quark action with nonlocal kernel induced by quark zero modes in the
instanton field (Section 3). The gauged version of the model
\cite{Birse95,ADoLT00,DoBr03} meets the symmetry properties with respect to
external gauge fields (Section 4), and corresponding vertices satisfy the
Ward-Takahashi identities. Below in Section 5 we demonstrate how the anomalous
structure $w_{L}$ is saturated within the instanton liquid model. We also
calculate the transversal invariant function $w_{T}$ at arbitrary $q$ and show
that within the instanton model at large $q^{2}$ there are no power
corrections to this structure. The nonperturbative corrections to $w_{T}$ at
large $q^{2}$ have exponentially decreasing behavior related to the short
distance properties of the instanton nonlocality in the QCD vacuum. We also
estimate the slope of transversal invariant function at zero virtuality.

When light quark current masses, $m_{f}$ $\left(  f=u,d\right)  ,$ are
switched on, additional OPE structures appear, with the leading one being of
dimension four $\sim m_{f}\bar{q}\,\sigma_{\alpha\beta}q$. Its matrix element
between vacuum and soft photon state is proportional to the quark condensate
magnetic susceptibility introduced in Ref.\thinspace\cite{Ioffe:1984ju}. Using
the expansion of the triangle amplitude in inverse powers of momentum transfer
squared we will derive an expression for the magnetic susceptibility in the
instanton model and find its momentum dependence (Section 6).

\section{The structure of $VA\widetilde{V}$ correlator}

We will employ a tensor decomposition of the $VVA$ triangle graph amplitude
suggested originally by Rosenberg \cite{Rosenberg:1963pp} for the general
kinematics of incoming momenta
\begin{align}
T_{\mu\nu\lambda}(q_{1},q_{2})  &  =A_{1}q_{1}^{\rho}\varepsilon_{\rho\mu
\nu\lambda}+A_{2}q_{2}^{\rho}\varepsilon_{\rho\mu\nu\lambda}+A_{3}q_{1}^{\nu
}q_{1}^{\rho}q_{2}^{\sigma}\varepsilon_{\rho\sigma\mu\lambda}\label{T}\\
&  +A_{4}q_{2}^{\nu}q_{1}^{\rho}q_{2}^{\sigma}\varepsilon_{\rho\sigma
\mu\lambda}+A_{5}q_{1}^{\mu}q_{1}^{\rho}q_{2}^{\sigma}\varepsilon_{\rho
\sigma\nu\lambda}+A_{6}q_{2}^{\mu}q_{1}^{\rho}q_{2}^{\sigma}\varepsilon
_{\rho\sigma\nu\lambda},\nonumber
\end{align}
where $q_{1}$ and $q_{2\text{ }}$are the vector field momenta with
corresponding Lorentz indices $\mu$ and $\nu$. The coefficients $A_{j}%
=A_{j}(q_{1},q_{2}),j=1,...6$ are the Lorentz invariant amplitudes. The vector
Ward identities provide a gauge invariant definition of the $A_{1}$ and
$A_{2}$ amplitudes in terms of finite amplitudes $A_{k},k=3,...,6,$
\begin{equation}
A_{1}=\left(  q_{1}q_{2}\right)  A_{3}+q_{2}^{2}A_{4},\qquad A_{2}=\left(
q_{1}q_{2}\right)  A_{6}+q_{1}^{2}A_{5}.\qquad\label{A12}%
\end{equation}

In the specific kinematics when one photon ($q_{2}\equiv q$) is virtual and
another one ($q_{1}$) represents the external electromagnetic field and can be
regarded as a real photon with the vanishingly small momentum $q_{1}$ only two
invariant functions survive in linear in small $q_{1}$ approximation
\cite{Kukhto92}. It is convenient to define longitudinal and transversal with
respect to axial current index amplitudes \cite{VainshPLB03}
\begin{equation}
w_{L}\left(  q^{2}\right)  =4\pi^{2}\widetilde{A}_{4}\left(  q^{2}\right)
,\qquad w_{T}\left(  q^{2}\right)  =4\pi^{2}\left(  \widetilde{A}_{4}\left(
q^{2}\right)  +\widetilde{A}_{6}\left(  q^{2}\right)  \right)  , \label{WLT}%
\end{equation}
where tilted amplitudes are $\widetilde{A}(q^{2})\equiv A\left(  q_{1}%
=0,q_{2}=q\right)  .$ In terms of $w$ invariant functions the $VA\widetilde
{V}$ amplitude becomes%

\begin{equation}
\widetilde{T}_{\mu\nu\lambda}(q_{1},q_{2})=\frac{1}{4\pi^{2}}\left[
w_{T}\left(  q_{2}^{2}q_{1}^{\rho}\varepsilon_{\rho\mu\nu\lambda}-q_{2}^{\nu
}q_{1}^{\rho}q_{2}^{\sigma}\varepsilon_{\rho\mu\sigma\lambda}+q_{2}^{\lambda
}q_{1}^{\rho}q_{2}^{\sigma}\varepsilon_{\rho\mu\sigma\nu}\right)  -w_{L}%
q_{2}^{\lambda}q_{1}^{\rho}q_{2}^{\sigma}\varepsilon_{\rho\mu\sigma\nu
}\right]  . \label{Tt}%
\end{equation}
Both structures are transversal with respect to vector current, $q_{2}^{\nu
}T_{\nu\lambda}=0$. As for the axial current, the first structure is
transversal with respect to $q_{2}^{\lambda}$ while the second is longitudinal
and thus anomalous.

The amplitude for the triangle diagrams (Fig.\thinspace\ref{fig:triangle}) can
be written as a correlator of the axial current $j_{\lambda}^{5}$ and two
vector currents $j_{\nu}$ and $\tilde{j}_{\mu}$
\begin{equation}
T_{\mu\nu\lambda}=-\int\mathrm{d}^{4}x\mathrm{d}^{4}y\,\mathrm{e}%
^{iqx-iky}\,\langle0|\,T\{j_{\nu}(x)\,\tilde{j}_{\mu}(y)\,j_{\lambda}%
^{5}(0)\}|0\rangle\,,
\end{equation}
where for light $u$ and $d$ quarks one has
\[
j_{\mu}=\bar{q}\,\gamma_{\mu}Qq,\qquad j_{\lambda}^{5}=\bar{q}\,\gamma
_{\lambda}\gamma_{5}\tau_{3}q\,,
\]
the quark field $q_{f}^{i}$ has color ($i$) and flavor ($f$) indices, the
charge matrix is $Q=\frac{1}{2}\left(  \frac{1}{3}+\tau_{3}\right)  $ and the
tilted current is for the soft momentum photon vertex.

In the local theory the one-loop result for the invariant functions $w_{T}$
and $w_{L}$ is\footnote{Here and below the small effects of isospin violation
is neglected, considering $m_{f}\equiv m_{u}=m_{d}$.}
\begin{equation}
w_{L}^{\mathrm{1-loop}}=2\,w_{T}^{\mathrm{1-loop}}=\frac{2N_{c}}{3}\int
_{0}^{1}\frac{\mathrm{d}\alpha\,\alpha(1-\alpha)}{\alpha(1-\alpha)q^{2}%
+m_{f}^{2}}\,, \label{wlt}%
\end{equation}
where the factor $N_{c}/3$ is due to color number and electric charge. The
analytical result for the triangle diagram with finite quark masses has been
obtained in \cite{Achasov:1992bu} by dispersion integral method. In the chiral
limit, $m_{f}=0$, one gets the result (\ref{WLTch}) (with additional factor
$N_{c}/3$).

When nonperturbative contributions to the triangle amplitude (Fig. 1c) are
taken into account it was shown in \cite{Knecht02} by using the OPE methods
that at large Euclidean $q^{2}$ the difference between the longitudinal and
transversal parts, $w_{LT}=w_{L}-2w_{T},$ starts in the chiral limit from
leading, $\sim1/q^{6},$ power behavior. The power terms are expected to
contribute only into the transversal function $w_{T}$. Below we demonstrate
that within the instanton liquid model in the chiral limit all allowed by OPE
power corrections to $w_{T}$ cancel each other and only exponentially
suppressed corrections remain.

\section{The instanton effective quark model}

To study nonperturbative effects in the triangle amplitude $\widetilde{T}%
_{\mu\nu\lambda}$ at low and high momenta one can use the framework of the
effective field model of QCD. In the low momenta domain the effect of the
nonperturbative structure of QCD vacuum become dominant. Since invention of
the QCD sum rule method based on the use of the standard OPE it is common to
parameterize the nonperturbative properties of the QCD vacuum by using
infinite towers of the vacuum expectation values of the quark-gluon operators.
From this point of view the nonlocal properties of the QCD vacuum result from
the partial resummation of the infinite series of power corrections, related
to vacuum averages of quark-gluon operators with growing dimension, and may be
conventionally described in terms of the nonlocal vacuum condensates
\cite{MikhRad92,DEM97}. This reconstruction leads effectively to nonlocal
modifications of the propagators and effective vertices of the quark and gluon fields.

The adequate model describing this general picture is the instanton liquid
model of QCD vacuum describing nonperturbative nonlocal interactions in terms
of the effective action \cite{ShSh}. Spontaneous breaking the chiral symmetry
and dynamical generation of a momentum-dependent quark mass are naturally
explained within the instanton liquid model. The nonsinglet and singlet $V$
and $A$ current-current correlators, the vector Adler function have been
calculated in \cite{DoBr03,ADpepanTop,ADprdG2} in the framework of the
effective chiral model with instanton-like nonlocal quark-quark interaction
\cite{ADoLT00}. In the same model the pion transition form factor normalized
by axial anomaly has been considered in \cite{AD02} for arbitrary photon virtualities.

We start with the nonlocal chirally invariant action which describes the
interaction of soft quark fields \cite{ADoLT00}
\begin{align}
S  &  =\int d^{4}x\ \overline{q}_{I}(x)\left[  i\gamma^{\mu}D_{\mu}%
-m_{f}\right]  q_{I}(x)+\frac{1}{2}G_{P}\int d^{4}X\int\prod_{n=1}^{4}%
d^{4}x_{n}\cdot\nonumber\\
&  \cdot\ f(x_{n})\left[  \overline{Q}(X-x_{1},X)\Gamma_{P}Q(X,X+x_{3}%
)\overline{Q}(X-x_{2},X)\Gamma_{P}Q(X,X+x_{4})\right]  , \label{Lint}%
\end{align}
where $D_{\mu}=\partial_{\mu}-iV_{\mu}\left(  x\right)  -i\gamma_{5}A_{\mu
}\left(  x\right)  $ and the matrix product $\left(  1\otimes1+i\gamma_{5}%
\tau^{a}\otimes i\gamma_{5}\tau^{a}\right)  $ provides the spin-flavor
structure of the interaction. In Eq.~(\ref{Lint}) $\overline{q}_{I}%
=(\overline{u},\overline{d})$ denotes the flavor doublet field of dynamically
generated quarks, $G_{P}$ is the four-quark coupling constant, and $\tau^{a}$
are the Pauli isospin matrices. The separable nonlocal kernel of the
interaction determined in terms of form factors $f(x)$ is motivated by
instanton model of QCD vacuum.

In order to make the nonlocal action gauge-invariant with respect to external
gauge fields $V_{\mu}^{a}(x)$ and $A_{\mu}^{a}(x)$, we define in (\ref{Lint})
the delocalized quark field, $Q(x),$ by using the Schwinger gauge phase
factor
\begin{equation}
Q(x,y)=P\exp\left\{  i\int_{x}^{y}dz_{\mu}\left[  V_{\mu}^{a}(z)+\gamma
_{5}A_{\mu}^{a}(z)\right]  T^{a}\right\}  q_{I}(y),\qquad\overline
{Q}(x,y)=Q^{\dagger}(x,y)\gamma^{0}, \label{Qxy}%
\end{equation}
where $P$ is the operator of ordering along the integration path, with $y$
denoting the position of the quark and $x$ being an arbitrary reference point.
The conserved vector and axial-vector currents have been derived earlier in
\cite{ADoLT00,DoBr03,ADprdG2}.

The dressed quark propagator, $S(p)$, is defined as
\begin{equation}
S^{-1}(p)=i\widehat{p}-M(p^{2}), \label{QuarkProp}%
\end{equation}
with the momentum-dependent quark mass found as the solution of the gap
equation
\begin{equation}
M(p^{2})=m_{f}+4G_{P}N_{f}N_{c}f^{2}(p^{2})\int\frac{d^{4}k}{\left(
2\pi\right)  ^{4}}f^{2}(k^{2})\frac{M(k^{2})}{k^{2}+M^{2}(k^{2})}.
\label{SDEq}%
\end{equation}
The formal solution is expressed as \cite{Birse95}
\begin{equation}
M(p^{2})=m_{f}+(M_{q}-m_{f})f^{2}(p^{2}),
\end{equation}
with constant $M_{q}\equiv M(0)$ determined dynamically from Eq.~(\ref{SDEq})
and the momentum dependent $f(p)$ is the normalized four-dimensional Fourier
transform of $f(x)$ given in the coordinate representation.

The nonlocal function $f(p)$ describes the momentum distribution of quarks in
the nonperturbative vacuum. Given nonlocality $f(p)$ the light quark
condensate in the chiral limit, $M(p)=M_{q}f^{2}(p)$, is expressed as
\begin{equation}
\left\langle 0\left|  \overline{q}q\right|  0\right\rangle =-N_{c}\int
\frac{d^{4}p}{4\pi^{4}}\frac{M(p^{2})}{p^{2}+M^{2}(p^{2})}. \label{QQI}%
\end{equation}
Its $n$-moment is proportional to the vacuum expectation value of the quark
condensate with the covariant derivative squared $D^{2}$ to the $n$th power
\begin{equation}
\left\langle 0\left|  \overline{q}D^{2n}q\right|  0\right\rangle =-N_{c}%
\int\frac{d^{4}p}{4\pi^{4}}p^{2n}\frac{M(p^{2})}{p^{2}+M^{2}(p^{2})}.
\label{20}%
\end{equation}
The $n$th moment of the quark condensate appears as a coefficient of Taylor
expansion of the nonlocal quark condensate defined as \cite{MikhRad92}
\begin{equation}
C(x)=\left\langle 0\left|  \overline{q}\left(  0\right)  P\exp\left[
i\int_{0}^{x}A_{\mu}\left(  z\right)  dz_{\mu}\right]  q\left(  x\right)
\right|  0\right\rangle \label{NLC}%
\end{equation}
with gluon Schwinger phase factor inserted for gauge invariance and the
integral is over the straight line path. Smoothness of $C(x)$ near $x^{2}=0$
leads to existence of the quark condensate moments in the \textit{l.h.s.} of
(\ref{20}) for any $n$. In order to make the integral in the \textit{r.h.s.}
of (\ref{20}) convergent the nonlocal function $f(p)$ for large arguments must
decrease faster than any inverse power of $p^{2}$, \textit{e.g.}, like some
exponential\footnote{Very similar arguments lead the author of
\cite{Teryaev04} to the conclusion that finiteness of all transverse momenta
moments of the quark distributions guarantees the exponential fall-of of the
cross sections.}
\begin{equation}
f\left(  p\right)  \sim\exp\left(  -\mathrm{const\cdot}p^{\alpha}\right)
,~\alpha>0\quad\mathrm{as\quad}p^{2}\rightarrow\infty. \label{ExpSup}%
\end{equation}

Note, that the operators entering the matrix elements in (\ref{20}) and
(\ref{NLC}) are constructed from the QCD quark and gluon fields. The
\textit{r.h.s.} of (\ref{20}) is the value of the matrix elements of
QCD\ defined operators calculated within the effective instanton model with
dynamical quark fields. Within the instanton model in the zero mode
approximation the function $f(p)$ depends on the gauge. It is implied
\cite{DEM97,DoLT98} that the \textit{r.h.s.} of (\ref{20}) corresponds to
calculations in the axial gauge for the quark effective field. It is selected
among other gauges because in this gauge the covariant derivatives become
ordinary ones: $D\rightarrow\partial,$ and the exponential in (\ref{NLC}) with
straight line path is reduced to unit. In particular it means that one uses
the quark zero modes in the instanton field given in the axial gauge when
define the gauge dependent dynamical quark mass. The axial gauge at large
momenta has exponentially decreasing behavior and all moments of the quark
condensate exist. In principle, to calculate the gauge invariant matrix
element corresponding to the of \textit{l.h.s.} of (\ref{20}) it is possible
to use the expression for the dynamical mass given in any gauge, but in that
case the factor $p^{2n}$ will be modified for more complicated weight function
providing invariance of the answer\footnote{If one would naively use the
dynamical quark mass corresponding to popular singular gauge then one finds
the problem with convergence of the integrals in (\ref{20}), because in this
gauge there is only powerlike asymptotics of $M\left(  p\right)  \sim p^{-6}$
at large $p^{2}.$}.

Furthermore, the large distance asymptotics of the instanton solution is also
modified by screening effects due to interaction of instanton field with
surrounding physical vacuum \cite{DEMM99,ADWB01}. To take into account these
effects and make numerics simpler we shell use for the nonlocal function the
Gaussian form
\begin{equation}
f(p)=\exp\left(  -p^{2}/\Lambda^{2}\right)  , \label{MassDyna}%
\end{equation}
where the parameter $\Lambda$ characterizes the nonlocality size and it is
proportional to the inverse average size of instanton in the QCD\ vacuum.

The important property of the dynamical mass (\ref{SDEq}) is that at low
virtualities passing through quark its mass is close to constituent mass,
while at large virtualities it goes to the current mass value. As we will see
in Sect. 5 this property is crucial in obtaining the anomaly at large momentum
transfer. The instanton liquid model can be viewed as an approximation of
large-$N_{c}$ QCD where the only new interaction terms, retained after
integration of the high frequency modes of the quark and gluon fields down to
a nonlocality scale $\Lambda$ at which spontaneous chiral symmetry breaking
occurs, are those which can be cast in the form of four-fermion ope\-ra\-tors
(\ref{Lint}). The parameters of the model are then the nonlocality scale
$\Lambda$ and the four-fermion coupling constant $G_{P}$.

\section{Conserved vector and axial-vector currents}

The quark-antiquark scattering matrix (Fig. 2) in pseudoscalar channel is
found from the Bethe-Salpeter equation as
\begin{equation}
\widehat{T}_{P}(q^{2})=\frac{G_{P}}{1-G_{P}J_{PP}(q^{2})}, \label{ScattMatr}%
\end{equation}
with the polarization operator being
\begin{equation}
J_{PP}(q^{2})=\int\frac{d^{4}k}{\left(  2\pi\right)  ^{4}}f^{2}\left(
k\right)  f^{2}\left(  k+q\right)  Tr\left[  S(k)\gamma_{5}S\left(
k+q\right)  \gamma_{5}\right]  . \label{J}%
\end{equation}

\begin{figure}[h]
\begin{center}
\includegraphics[width=10cm]{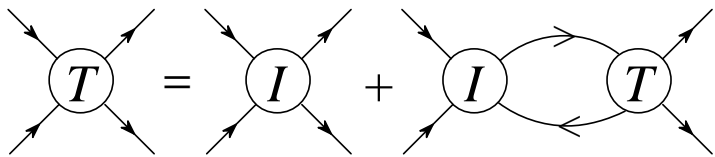}
\end{center}
\caption{{Diagrammatic representation of the Bethe-Salpeter equation for the
quark-quark scattering matrix, $T$, with nonlocal instanton kernel, }$I$.}%
\label{BSTfig}%
\end{figure}

The position of pion state is determined as the pole of the scattering matrix
\begin{equation}
\left.  \det(1-G_{P}J_{PP}(q^{2}))\right|  _{q^{2}=-m_{\pi}^{2}}=0.
\label{PoleEq}%
\end{equation}
The quark-pion vertex found from the residue of the scattering matrix is
$\left(  k^{\prime}=k+q\right)  $
\begin{equation}
\Gamma_{\pi}^{a}\left(  k,k^{\prime}\right)  =g_{\pi qq}i\gamma_{5}%
f(k)f(k^{\prime})\tau^{a}\quad\nonumber
\end{equation}
with the quark-pion coupling found from
\begin{equation}
g_{\pi q}^{-2}=-\left.  \frac{dJ_{ii}\left(  q^{2}\right)  }{dq^{2}}\right|
_{q^{2}=-m_{\pi}^{2}}, \label{gM}%
\end{equation}
where $m_{\mathrm{\pi}}$ is physical mass of the $\pi$-meson. The quark-pion
coupling, $g_{\pi q}^{2}$, and the pion decay constant, $f_{\pi}$, are
connected by the Goldberger-Treiman relation, $g_{\pi}=M_{q}/f_{\pi},$ which
is verified to be valid in the nonlocal model \cite{Birse95}, as requested by
the chiral symmetry.

\begin{figure}[h]
\begin{center}
\includegraphics[width=10cm]{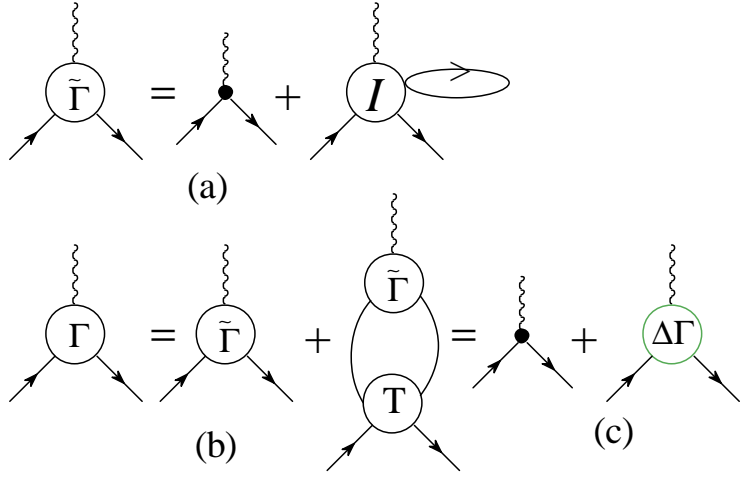}
\end{center}
\caption{{Diagrammatic representation of the bare (a) and full (b)
quark-current vertices. Diagram (c) shows separation of local (fat dot) and
nonlocal parts of the full vertex.}}%
\label{w5}%
\end{figure}

The vector vertex following from the model (\ref{Lint}) is (Fig. \ref{w5}a)
\begin{equation}
\Gamma_{\mu}(k,k^{\prime})=\gamma_{\mu}+(k+k^{\prime})_{\mu}M^{(1)}%
(k,k^{\prime}), \label{GV}%
\end{equation}
where $M^{(1)}(k,k^{\prime})$ is the finite-difference derivative of the
dynamical quark mass, $q$ is the momentum corresponding to the current, and
$k$ $(k^{\prime})$ is the incoming (outgoing) momentum of the quark,
$k^{\prime}=k+q$. The finite-difference derivative of an arbitrary function
$F$ is defined as
\begin{equation}
F^{(1)}(k,k^{\prime})=\frac{F(k^{\prime})-F(k)}{k^{\prime2}-k^{2}}.
\label{FDD}%
\end{equation}

The full axial vertex corresponding to the conserved axial-vector current is
obtained after resummation of quark-loop chain that results in appearance of
term proportional to the pion propagator \cite{ADoLT00} (Fig. \ref{w5}b)
\begin{align}
\Gamma_{\mu}^{5}(k,k^{\prime})  &  =\gamma_{\mu}\gamma_{5}+2\gamma_{5}%
\frac{q_{\mu}}{q^{2}}f(k)f(k^{\prime})\left[  J_{AP}\left(  0\right)
-\frac{m_{f}G_{P}J_{P}\left(  q^{2}\right)  }{1-G_{P}J_{PP}\left(
q^{2}\right)  }\right] \label{GAtot}\\
&  +(k+k^{\prime})_{\mu}J_{AP}\left(  0\right)  \frac{\left(  f(k^{\prime
})-f\left(  k\right)  \right)  ^{2}}{k^{\prime2}-k^{2}},\nonumber
\end{align}
where we have introduced the notations
\begin{equation}
J_{P}(q^{2})=\int\frac{d^{4}k}{\left(  2\pi\right)  ^{4}}f\left(  k\right)
f\left(  k+q\right)  Tr\left[  S(k)\gamma_{5}S\left(  k+q\right)  \gamma
_{5}\right]  . \label{JP}%
\end{equation}%
\begin{equation}
J_{AP}(q^{2})=4N_{c}N_{f}\int\frac{d^{4}l}{\left(  2\pi\right)  ^{4}}%
\frac{M\left(  l\right)  }{D\left(  l\right)  }\sqrt{M\left(  l+q\right)
M\left(  l\right)  }. \label{JAP}%
\end{equation}
The axial-vector vertex has a pole at
\begin{equation}
q^{2}=-m_{\pi}^{2}=m_{c}\left\langle \overline{q}q\right\rangle /f_{\pi}^{2}%
\end{equation}
where the Goldberger-Treiman relation and definition of the quark condensate
have been used. The pole is related to the denominator $1-G_{P}J_{PP}\left(
q^{2}\right)  $ in Eq. (\ref{GAtot}), while $q^{2}$ in denominator is
compensated by zero from square brackets in the limit $q^{2}\rightarrow0.$
This compensation follows from expansion of $J(q^{2})$ functions near zero
\begin{align*}
J_{PP}(q^{2})  &  =G_{P}^{-1}+m_{c}\left\langle \overline{q}q\right\rangle
M^{-2}\left(  0\right)  -q^{2}g_{\pi q}^{-2}+O\left(  q^{4}\right)  ,\qquad\\
J_{AP}(q^{2}  &  =0)=M\left(  0\right)  ,\qquad J_{P}(q^{2}=0)=\left\langle
\overline{q}q\right\rangle M^{-1}\left(  0\right)  .
\end{align*}
In the chiral limit $m_{f}=0$ the second structure in square brackets in Eq.
(\ref{GAtot}) disappears and the pole moves to zero.

The parameters of the model are fixed in a way typical for effective
low-energy quark models. One usually fits the pion decay constant, $f_{\pi}$,
to its experimental value, which in the chiral limit reduces to $86$
\textrm{MeV} \cite{LeutG}. In the instanton model the constant, $f_{\pi}$, is
determined by
\begin{equation}
f_{\pi}^{2}=\frac{N_{c}}{4\pi^{2}}\int\limits_{0}^{\infty}du\ u\frac
{M^{2}(u)-uM(u)M^{\prime}(u)+u^{2}M^{\prime}(u)^{2}}{D^{2}\left(  u\right)  },
\label{Fpi2_M}%
\end{equation}
where here and below $u=k^{2},$ primes mean derivatives with respect to $u$:
$M^{\prime}(u)=dM(u)/du$, \textit{etc.}, and
\[
D\left(  k^{2}\right)  =k^{2}+M(k)^{2}.
\]

One gets the values of the model parameters \cite{ADprdG2}
\begin{equation}
M_{q}=0.24~\mathrm{GeV,}\qquad\Lambda_{P}=1.11~\mathrm{GeV,\quad}%
G_{P}=27.4~\mathrm{GeV}^{-2}. \label{G's}%
\end{equation}

\section{$VA\widetilde{V}$ correlator within the instanton liquid model}

Our goal is to obtain the nondiagonal correlator of vector current and
nonsinglet axial-vector current in the external electromagnetic field
($VA\widetilde{V}$) by using the effective instanton-like model (\ref{Lint}).
In this model the $VA\widetilde{V}$ correlator is defined by (Fig. 4a)
\begin{align}
\widetilde{T}_{\mu\nu\lambda}(q_{1},q_{2})  &  =-2N_{c}\int\frac{d^{4}%
k}{\left(  2\pi\right)  ^{4}}\cdot\label{Tncqm}\\
\cdot &  Tr\left[  \Gamma_{\mu}\left(  k+q_{1},k\right)  S\left(
k+q_{1}\right)  \Gamma_{\lambda}^{5}\left(  k+q_{1},k-q_{2}\right)  S\left(
k-q_{2}\right)  \Gamma_{\nu}\left(  k,k-q_{2}\right)  S\left(  k\right)
\right]  ,\nonumber
\end{align}
where the quark propagator, the vector and the axial-vector vertices are given
by (\ref{QuarkProp}), (\ref{GV}) and (\ref{GAtot}), respectively. The
structure of the vector vertices guarantees that the amplitude is transversal
with respect to vector indices
\begin{equation}
\widetilde{T}_{\mu\nu\lambda}(q_{1},q_{2})q_{1}^{\mu}=\widetilde{T}_{\mu
\nu\lambda}(q_{1},q_{2})q_{2}^{\nu}=0
\end{equation}
and the Lorentz structure of the amplitude is given by (\ref{Tt}).

It is convenient to express Eq. (\ref{Tncqm}) as a sum of the contribution
where all vertices are local (Fig. 4b), and the rest contribution containing
nonlocal parts of the vertices (Fig. 3c). Further results in this section will
concern the chiral limit.

\begin{figure}[h]
\begin{center}
\includegraphics[height=4cm]{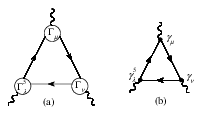}
\end{center}
\caption{{Diagrammatic representation of the triangle diagram in the instanton
model with dressed quark lines and full quark-current vertices (a); and part
of the diagram when all vertices are local one (b). }}%
\label{w6}%
\end{figure}

The contributions of diagram 4b to the invariant functions at space-like
momentum transfer, $q^{2}\equiv q_{2}^{2}$, are given by
\begin{align}
\widetilde{A}_{4}^{\left(  L\right)  }\left(  q^{2}\right)   &  =\frac{N_{c}%
}{9q^{2}}\int\frac{d^{4}k}{\pi^{4}}\frac{1}{D_{+}^{2}D_{-}}\left[
k^{2}-4\frac{\left(  kq\right)  ^{2}}{q^{2}}+3\left(  kq\right)  \right]
,\label{A46a}\\
\qquad\widetilde{A}_{6}^{\left(  L\right)  }\left(  q^{2}\right)   &
=-\frac{1}{2}\widetilde{A}_{4}^{\left(  L\right)  }\left(  q^{2}\right)  .
\label{A6a}%
\end{align}
where the notations used here and below are
%
%
%
%
%
%
%
%
%
%
%
%
%
%
%
%
%
%
%
\[
k_{+}=k,\qquad k_{-}=k-q,\qquad k_{\perp}^{2}=k_{+}k_{-}-\frac{\left(
k_{+}q\right)  \left(  k_{-}q\right)  }{q^{2}},
\]%
\[
D_{\pm}=D(k_{\pm}^{2}),\qquad M_{\pm}=M(k_{\pm}^{2}),\ \ \ \ f_{\pm}=f(k_{\pm
}^{2}).\ \ \ \
\]

At large $q^{2}$ one has an expansion
\begin{equation}
\widetilde{A}_{4}^{\left(  L\right)  }\left(  q^{2}\rightarrow\infty\right)
=\frac{N_{c}}{6\pi^{2}}\left(  \frac{1}{q^{2}}+\frac{a_{\left(  4\right)
}^{\left(  L\right)  }}{q^{4}}+\frac{a_{\left(  6\right)  }^{\left(  L\right)
}}{q^{6}}+O\left(  q^{-8}\right)  \right)  , \label{A4as}%
\end{equation}
with coefficients given by
\begin{equation}
a_{\left(  4\right)  }^{\left(  L\right)  }=-\int_{0}^{\infty}du\frac
{M^{2}\left(  u\right)  }{D^{2}\left(  u\right)  }\left(  2u+M^{2}\left(
u\right)  \right)  ,\qquad a_{\left(  6\right)  }^{\left(  L\right)  }%
=-\frac{2}{3}\int_{0}^{\infty}du\frac{uM^{2}\left(  u\right)  \left(
u+2M^{2}\left(  u\right)  \right)  }{D^{2}\left(  u\right)  }. \label{A4aAs}%
\end{equation}
It is clear that the contribution (\ref{A46a}) saturate the anomaly at large
$q^{2}$. The reason is that the leading asymptotics of (\ref{A46a}) is given
by the configuration where the large momentum is passing through all quark
lines. Then the dynamical quark mass $M(k)$ reduces to zero and the asymptotic
limit of triangle diagram with dynamical quarks and local vertices coincides
with the standard triangle amplitude with massless quarks and, thus, it is
independent of the model.

The contribution to the form factors when the nonlocal parts of the vector and
axial-vector vertices are taken into account is given by
\begin{align}
\widetilde{A}_{4}^{\left(  NL\right)  }\left(  q^{2}\right)   &  =\frac{N_{c}%
}{3q^{2}}\int\frac{d^{4}k}{\pi^{4}}\frac{1}{D_{+}^{2}D_{-}}\left\{
M_{+}\left[  M_{+}-\frac{4}{3}M_{+}^{\prime}k_{\perp}^{2}\right]  -\right.
\nonumber\\
&  \left.  -M^{2(1)}(k_{+},k_{-})\left(  2\frac{\left(  kq\right)  ^{2}}%
{q^{2}}-\left(  kq\right)  \right)  \right\}  . \label{A4bt}%
\end{align}
One has for the leading terms of large $q^{2}$ asymptotics
\begin{equation}
\widetilde{A}_{4}^{\left(  NL\right)  }\left(  q^{2}\rightarrow\infty\right)
=\frac{N_{c}}{6\pi^{2}}\left(  \frac{a_{\left(  4\right)  }^{\left(
NL\right)  }}{q^{4}}+\frac{a_{\left(  6\right)  }^{\left(  NL\right)  }}%
{q^{6}}+O\left(  q^{-8}\right)  \right)  , \label{A4bAs}%
\end{equation}
with coefficients given by
\begin{align}
a_{\left(  4\right)  }^{\left(  NL\right)  }  &  =2\int_{0}^{\infty}%
du\frac{uM\left(  u\right)  }{D^{2}\left(  u\right)  }\left(  M\left(
u\right)  -uM^{\prime}\left(  u\right)  \right)  ,\qquad\label{a4bc}\\
a_{\left(  6\right)  }^{\left(  NL\right)  }  &  =\frac{2}{3}\int_{0}^{\infty
}du\frac{u^{3}M\left(  u\right)  M^{\prime}\left(  u\right)  }{D^{2}\left(
u\right)  }.\nonumber
\end{align}
In sum of two contributions both power corrections with coefficients
$a_{\left(  4\right)  }$ and $a_{\left(  6\right)  }$ are canceled. To prove
cancellation for the $a_{\left(  4\right)  }$ coefficient one needs to use
integration by parts.

Summing analytically the local (\ref{A46a}) and nonlocal (\ref{A4bt}) parts
provides us with the result required by the axial anomaly
\begin{equation}
w_{L}(q^{2})=4\pi^{2}\widetilde{A}_{4}\left(  q^{2}\right)  =\frac{2N_{c}}%
{3}\frac{1}{q^{2}}. \label{A4Tot}%
\end{equation}
Fig. 5 illustrates how different contributions saturate the anomaly. Note,
that at zero virtuality the saturation of anomaly follows from anomalous
diagram of pion decay in two photons. This part is due to the triangle diagram
involving nonlocal part of the axial vertex and local parts of the photon
vertices. The result (\ref{A4Tot}) is in agreement with the statement about
absence of nonperturbative corrections to longitudinal invariant function
following from the 't Hooft duality arguments. Earlier this consistency has
also been demonstrated within the QCD sum rules \cite{NestRad83,Pivovarov91}
and within dispersion method \cite{TerHor95} considering the lowest orders of
expansion of the triangle diagram in condensates.

For $\widetilde{A}_{6}^{\left(  NL\right)  }\left(  q^{2}\right)  $ invariant
function one gets
\begin{align}
\widetilde{A}_{6}^{\left(  NL\right)  }\left(  q^{2}\right)   &  =-\frac
{N_{c}}{6q^{2}}\int\frac{d^{4}k}{\pi^{4}}\frac{1}{D_{+}^{2}D_{-}}\left\{
\left(  M_{+}+M_{-}\right)  \left[  M_{+}-\frac{\left(  kq\right)  }{q^{2}%
}\left(  M_{+}-M_{-}\right)  \right]  -\right.  \label{A4bc}\\
&  \left.  -\frac{2}{3}M_{+}^{\prime}\left[  2k_{\perp}^{2}M_{+}-M_{-}%
\frac{q^{2}}{k_{+}^{2}-k_{-}^{2}}\left(  k^{2}+2\frac{\left(  kq\right)  ^{2}%
}{q^{2}}-6\left(  kq\right)  \frac{k^{2}}{q^{2}}\right)  \right]  \right\}
+\nonumber\\
&  +\frac{2N_{c}}{9q^{2}}\int\frac{d^{4}k}{\pi^{4}}\frac{\sqrt{M_{+}M_{-}}%
}{D_{+}^{2}D_{-}}\frac{k_{\perp}^{2}}{k_{+}^{2}-k_{-}^{2}}\left[  M_{+}%
-M_{-}-2M_{+}^{\prime}\left(  kq\right)  \right]  .\nonumber
\end{align}

Then, let us consider the combination of invariant functions which show up
nonperturbative dynamics
\begin{equation}
w_{LT}(q^{2})\equiv w_{L}(q^{2})-2w_{T}(q^{2})=-4\pi^{2}\left[  \widetilde
{A}_{4}\left(  q^{2}\right)  +2\widetilde{A}_{6}\left(  q^{2}\right)  \right]
.\label{WLTtot}%
\end{equation}
From (\ref{A6a}) we see that the contribution to $w_{LT}$ from the triangle
diagram 4b with local vertices is absent. In sum of $\widetilde{A}_{4}\left(
q^{2}\right)  $ and $\widetilde{A}_{6}\left(  q^{2}\right)  $ a number of
cancellations takes place and the final result is quite simple%
\begin{align}
w_{LT}\left(  q^{2}\right)   &  =\frac{4N_{c}}{3q^{2}}\int\frac{d^{4}k}%
{\pi^{2}}\frac{\sqrt{M_{-}}}{D_{+}^{2}D_{-}}\left\{  \sqrt{M_{-}}\left[
M_{+}-\frac{2}{3}M_{+}^{\prime}\left(  k^{2}+2\frac{\left(  kq\right)  ^{2}%
}{q^{2}}\right)  \right]  -\right.  \nonumber\\
&  \left.  -\frac{4}{3}k_{\perp}^{2}\left[  \sqrt{M_{+}}M^{(1)}(k_{+}%
,k_{-})-2\left(  kq\right)  M_{+}^{\prime}\sqrt{M}^{\left(  1\right)  }%
(k_{+},k_{-})\right]  \right\}  .\label{WLTf}%
\end{align}
The behavior of $w_{LT}(q^{2})$ is presented in Fig. 6.

\begin{figure}[h]
\hspace*{1cm} \begin{minipage}{7cm}
\vspace*{0.5cm} \epsfxsize=6cm \epsfysize=5cm \centerline{\epsfbox{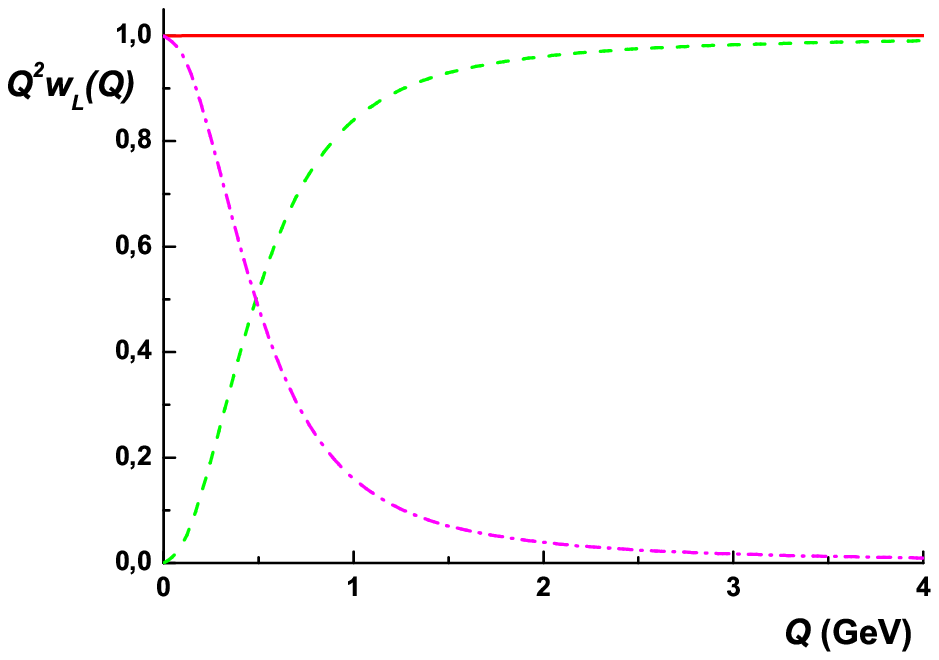}}
\caption[dummy0]{ Normalized $w_L$
invariant function constrained by ABJ anomaly from triangle diagram Fig. 4a (solid line)
and different contributions to
it: from local part, Fig. 4b, (dashed line), and from the nonlocal part
(dash-dotted line).
\label{WLfig} }
\end{minipage}\hspace*{0.5cm} \begin{minipage}{7cm}
\vspace*{0.5cm} \epsfxsize=6cm \epsfysize=5cm \centerline{\epsfbox
{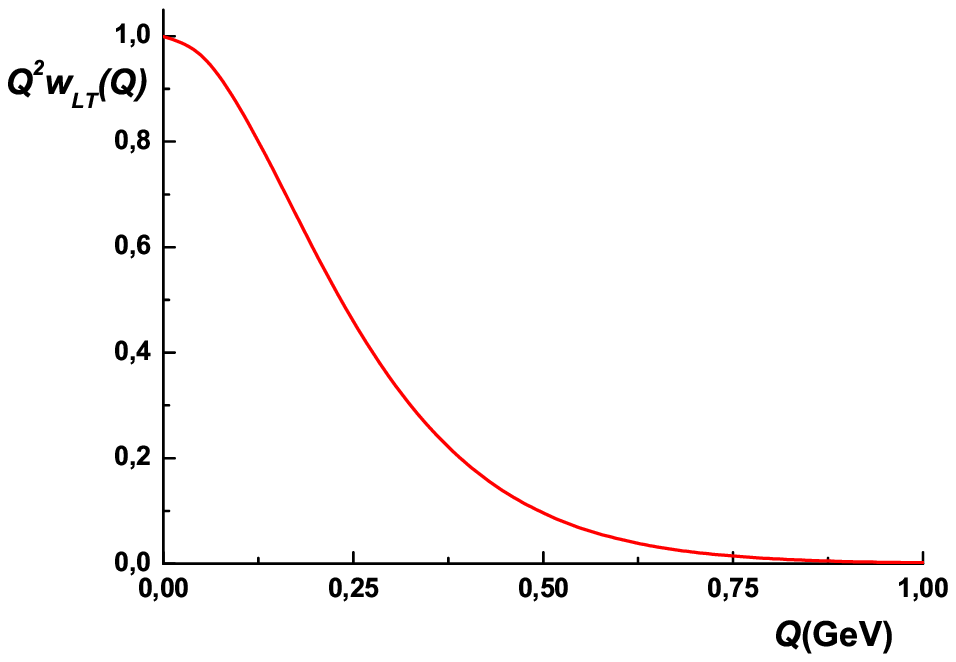}}
\caption[dummy0]{ Normalized $w_{LT}$
invariant function  versus
$Q$ predicted by the instanton model from the diagram Fig. 4a.
\label{WLTfig} }
\end{minipage}
\end{figure}

In the above expression the integrand is proportional to the product of
nonlocal form factors $f\left(  k_{+}\right)  f\left(  k_{-}\right)  $
depending on quark momenta passing through different quark lines. Then, it
becomes evident that the large $q^{2}$ asymptotics of the integral is governed
by the asymptotics of the nonlocal form factor $f\left(  q\right)  $ which is
exponentially suppressed (\ref{ExpSup}). Thus, within the instanton model the
distinction between longitudinal and transversal parts is exponentially
suppressed at large $q^{2}$ and all allowed by OPE power corrections are
canceled each other. Recently, it was proven that the relation
\begin{equation}
w_{LT}[m_{f}=0]=0, \label{wtwl}%
\end{equation}
which holds at the one-loop level gets no perturbative corrections from gluon
exchanges \cite{VainshPLB03}. The instanton liquid model indicates that it may
be possible that due to anomaly this relation is violated at large $q^{2}$
only exponentially.

We also find numerical values of the slope of invariant function $w_{LT}$ at
zero virtuality
\begin{equation}
\left.  \frac{\partial w_{LT}(q^{2})}{\partial q^{2}}\right|  _{q^{2}%
=0}\left(  \mu_{\mathrm{Inst}}\right)  =-3.8~\mathrm{GeV}^{-2},
\end{equation}
and its width
\begin{equation}
\lambda_{LT}^{2}\equiv\int uw_{LT}\left(  u\right)  du\cdot\left(  \int
w_{LT}\left(  u\right)  du\right)  ^{-1}=0.54~\mathrm{GeV}^{2}.
\end{equation}

\section{Magnetic susceptibility of quark condensate}

In this section we consider the leading power corrections to $w_{L}$ and
$w_{T}$ resulting from inclusion of current quark mass, $m_{f}$, into
consideration. An appearance of this kind of power corrections is already
clear from perturbative expression (\ref{wlt}). In OPE the leading, by
dimension, correction to the invariant functions $w_{T,L}(q^{2})$ is
\begin{equation}
\Delta w_{L}=2\,\Delta w_{T}=\frac{4m_{f}\kappa_{f}}{3q^{4}}\,, \label{dW(3)}%
\end{equation}
where $\kappa_{f}$ are the matrix element of dimension 3 operators
\begin{equation}
\mathcal{O}_{f}^{\alpha\beta}=-i\,\bar{q}_{f}\,\sigma^{\alpha\beta}\gamma
_{5}\,q^{f}\,, \label{fop}%
\end{equation}
between the soft photon and vacuum states. Proportionality to $m_{f}$ in
(\ref{dW(3)}) is in correspondence with chirality arguments.

In perturbation theory the matrix element $\kappa_{f}$ of the chirality-flip
operator $O_{f}$ is proportional to $m_{f}$. Nonperturbatively, however, due
to spontaneous breaking of the chiral symmetry $\kappa_{f}$ does not vanish at
$m_{f}\!=\!0\,$. It is convenient to introduce the magnetic susceptibility
$\chi_{m}$ normalized by the quark condensate \cite{Ioffe:1984ju}
\[
\kappa_{f}=-4\pi^{2}\,\langle\bar{q}q\rangle\,\chi_{m}\,.
\]
This representation emphasizes that magnetic susceptibility for the
nondiagonal vector--axial-vector vector correlator in the external
electromagnetic field plays the similar role as the quark condensate for the
diagonal correlators of vector and axial-vector currents.

In the instanton model the $VA\widetilde{V}$ correlator is given by
(\ref{Tncqm}) with the quark propagator, the vector and the axial-vector
vertices defined by (\ref{QuarkProp}), (\ref{GV}) and (\ref{GAtot}), with the
quark mass, $m_{f}$, being included. Keeping in the calculation only linear in
the current quark mass terms one finds at large $q^{2}$ the correction at
twist 4 level (\ref{dW(3)}) for the contribution of diagram Fig. 4b%

\begin{equation}
\Delta\widetilde{A}_{4}^{\left(  L\right)  }\left(  q^{2}\rightarrow
\infty\right)  =-\frac{1}{q^{4}}\frac{2m_{f}}{3}\frac{N_{c}}{\pi^{2}}\int
du\frac{u^{2}M\left(  u\right)  }{D^{3}\left(  u\right)  }, \label{A4Lm}%
\end{equation}
and from nonlocal part%
\begin{equation}
\Delta\widetilde{A}_{4}^{\left(  NL\right)  }\left(  q^{2}\rightarrow
\infty\right)  =-\frac{1}{q^{4}}\frac{m_{f}}{3}\frac{N_{c}}{\pi^{2}}\int
du\frac{uM\left(  u\right)  }{D^{3}\left(  u\right)  }\left[  -u+3M^{2}\left(
u\right)  -4uM\left(  u\right)  M^{\prime}\left(  u\right)  \right]  .
\label{A4NLm}%
\end{equation}
The leading asymptotics linear in current mass for the invariant function
$\widetilde{A}_{6}$ is given by the relation%
\begin{equation}
\Delta\widetilde{A}_{6}\left(  q^{2}\rightarrow\infty\right)  =-\frac{1}%
{2}\Delta\widetilde{A}_{4}\left(  q^{2}\rightarrow\infty\right)  \label{A6m}%
\end{equation}
which is in accordance with OPE (\ref{dW(3)}).

Then, summing up contributions (\ref{A4Lm}) and (\ref{A4NLm}) and comparing
the result at large $q^{2}$ with OPE one gets the magnetic susceptibility in
the form
\begin{equation}
\chi_{m}\left(  \mu_{\mathrm{Inst}}\right)  =-\frac{1}{\left\langle 0\left|
\overline{q}q\right|  0\right\rangle }\frac{N_{c}}{4\pi^{2}}\int
du\frac{u\left(  M\left(  u\right)  -uM^{\prime}\left(  u\right)  \right)
}{D^{2}\left(  u\right)  }, \label{ChiInst}%
\end{equation}
where the quark condensate is defined in (\ref{QQI}).

Alternatively, to get (\ref{ChiInst}) we may simply calculate the matrix
element of $\mathcal{O}_{f}^{\alpha\beta}$ (\ref{fop}) between vacuum and one
real photon state and use Eq. (\ref{GV}) for the quark-photon vertex. In this
way it is easy to show that the result (\ref{ChiInst}) stays unchanged when
one includes the vector meson degrees of freedom. Indeed, in the extended
model \cite{ADprdG2,DoMKR} the vector vertex gets contribution from vector
$\rho$ and $\omega$ mesons in the form%
\begin{equation}
\Delta\Gamma_{\mu}^{a}(p,p^{\prime})=\left(  g_{\mu\nu}-\frac{q^{\mu}q^{\nu}%
}{q^{2}}\right)  \gamma_{\nu}T^{a}\frac{G_{V}f^{V}\left(  p\right)
f^{V}\left(  p^{\prime}\right)  }{1-G_{V}J_{V}^{T}\left(  q^{2}\right)  }%
B_{V}\left(  q^{2}\right)  , \label{GVdress}%
\end{equation}
where $T^{a}$\ is a flavor matrix, $f^{V}\left(  p\right)  $, $J_{V}%
^{T}\left(  q^{2}\right)  ,$ $G_{V}$ are the nonlocal form factor, the
polarization operator and four-quark coupling in the vector channel,
correspondingly. Due to conservation of the vector current one has
$B_{V}\left(  q^{2}=0\right)  =0$ and thus there is no contribution to the
magnetic susceptibility.

It is easy also to derive the momentum dependence of the magnetic
susceptibility%
\begin{align}
\chi_{m}\left(  q\right)   &  =-\frac{N_{c}}{\left\langle 0\left|
\overline{q}q\right|  0\right\rangle }\int\frac{d^{4}k}{4\pi^{4}}\frac
{1}{D_{+}D_{-}}\left\{  \left[  M_{+}-\frac{kq}{q^{2}}\left(  M_{+}%
-M_{-}\right)  \right]  \left(  1+B_{V}\left(  q^{2}\right)  f_{+}^{V}%
f_{-}^{V}\right)  -\right. \label{ChiQ}\\
&  \left.  -\frac{2}{3}k_{\perp}^{2}M^{(1)}(k_{+},k_{-})\right\}  ,\nonumber
\end{align}
presented in Fig. 7. At large $q$ the integral in (\ref{ChiQ}) is proportional
to the quark condensate providing the asymptotic result%
\begin{equation}
\chi_{m}\left(  q\rightarrow\infty\right)  =\frac{2}{q^{2}}. \label{ChiAs}%
\end{equation}

\begin{figure}[th]
\begin{center}
\includegraphics[width=6cm]{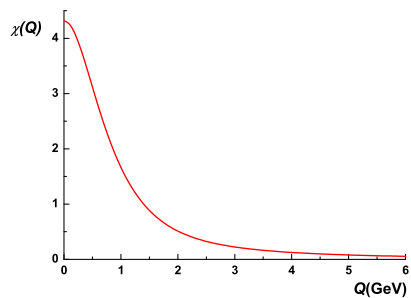}
\end{center}
\caption{The momentum dependence of the magnetic susceptibility of quark
condensate.}%
\label{ChiF}%
\end{figure}

Recently, Eq. (\ref{ChiInst}) has been obtained by more complicated way in
\cite{MusaPLB05}. Also note that the instanton model does not support use of
the pion dominance for estimate of the magnetic susceptibility as it was
attempted in \cite{VainshPLB03}. The reason is that the pion pole in the axial
vertex (\ref{GAtot}) is accompanied by the exponentially suppressed residue
$J_{P}\left(  q^{2}\right)  $. Thus, it does not contribute to the twist 4 coefficient.

Given model parameters (\ref{G's}) one finds numerical values for the quark
condensate and the magnetic susceptibility
\begin{equation}
\left\langle 0\left|  \overline{q}q\right|  0\right\rangle \left(
\mu_{\mathrm{Inst}}\right)  =-\left(  214~\mathrm{GeV}\right)  ^{3},\qquad
\chi_{m}\left(  \mu_{\mathrm{Inst}}\right)  =4.32~\mathrm{GeV}^{-2}\mathrm{,}%
\end{equation}
where $\mu_{\mathrm{Inst}}$ is the normalization scale typical for instanton
fluctuations. To leading-logarithmic accuracy scale dependence of these values
is predicted by QCD as%
\begin{equation}
\left\langle 0\left|  \overline{q}q\right|  0\right\rangle \left(  \mu\right)
=L^{-\gamma_{\overline{q}q}/b}\left\langle 0\left|  \overline{q}q\right|
0\right\rangle \left(  \mu_{0}\right)  ,\qquad\chi_{m}\left(  \mu\right)
=L^{-\left(  \gamma_{0}-\gamma_{\overline{q}q}\right)  /b}\chi_{m}\left(
\mu_{0}\right)  ,
\end{equation}
where $L=\alpha_{s}\left(  \mu\right)  /\alpha_{s}\left(  \mu_{0}\right)  $,
$b=\left(  11N_{c}-2n_{f}\right)  /3$, $\gamma_{\overline{q}q}=-3C_{F}$ is the
anomalous dimension of the quark condensate, $\gamma_{0}=C_{F}$ is the
anomalous dimension of the chiral-odd local operator of leading-twist,
$C_{F}=4/3$. We may fix the normalization scale of the model by comparing the
value of the condensate with that found in QCD sum rule at some standard
normalization point: $\left\langle 0\left|  \overline{q}q\right|
0\right\rangle \left(  \mu_{0}=1~\mathrm{GeV}\right)  =-\left(
240~\mathrm{GeV}\right)  ^{3}$. Then one finds $L=2.17$ that corresponds to
the normalization point $\mu_{\mathrm{Inst}}\approx0.5~\mathrm{GeV}$, with the
QCD constant for three flavors being $\Lambda_{\mathrm{QCD}}^{\left(
n_{f}=3\right)  }=296$ MeV. The rescaled magnetic susceptibility calculated in
the model will be%
\begin{equation}
\chi_{m}\left(  \mu_{\mathrm{Inst}}=1~\mathrm{GeV}\right)  =2.73~\mathrm{GeV}%
^{-2},
\end{equation}
which is in rather good agreement with the latest numerical value of $\chi
_{m}$ obtained with the QCD sum rule fit \cite{BBK}: $\chi_{m}\left(
\mu_{\mathrm{SR}}=1~\mathrm{GeV}\right)  =\left(  3.15\pm0.3\right)
~\mathrm{GeV}^{-2}$. A phenomenology of hard exclusive processes sensitive to
the magnetic susceptibility $\chi_{m}\,$, see \cite{BBK}, will possibly help
to fix its value.

\section{Conclusions}

In the framework of the instanton liquid model we calculated for arbitrary
momenta transfer the nondiagonal correlator of the vector and nonsinglet
axial-vector currents in the background of a soft vector field. In this case
we find that at large momenta the nonperturbative power corrections are absent
in the chiral limit for the transversal part $w_{T}$ of triangle diagram. The
transversal part is corrected only by exponentially small terms which reflects
the nonlocal structure of QCD vacuum. Within the instanton model the
saturation of the anomalous, longitudinal $w_{L}$ structure is demonstrated
explicitly. Using the instanton liquid model we also derive an expression for
the quark condensate magnetic susceptibility and its momentum dependence.

The author is grateful to A. P. Bakulev, N. I. Kochelev, P. Kroll, S. V.
Mikhailov, A. A. Pivovarov, O. V. Teryaev for helpful discussions on the
subject of the present work. The author also thanks for partial support from
the Russian Foundation for Basic Research projects nos. 03-02-17291,
04-02-16445 and the Heisenberg--Landau program.

\end{document}